\documentstyle[12pt,epsfig]{article}
\begin{document}
\def\beq{\begin{equation}}
\def\eeq{\end{equation}}
\def\bea{\begin{eqnarray}}
\def\eea{\end{eqnarray}}
\def\ve{\vert}
\def\vel{\left|}
\def\ver{\right|}
\def\nnb{\nonumber}
\def\ga{\left(}
\def\dr{\right)}
\def\aga{\left\{}
\def\adr{\right\}}
\def\rar{\rightarrow}
\def\nnb{\nonumber}
\def\la{\langle}
\def\ra{\rangle}
\def\ba{\begin{array}}
\def\ea{\end{array}}
\def\tep{$B \rar K \ell^+ \ell^-$}
\def\tepm{$B \rar K \mu^+ \mu^-$}
\def\tept{$B \rar K \tau^+ \tau^-$}
\def\ds{\displaystyle}

\baselineskip 20pt        

\vspace*{2.8cm}

\begin{center}
{\Large \bf
Exclusive $B \rar M \ell^+ \ell^-~(M=\pi,~K,~\rho,~K^*)$ Decays and
Determinations of $\vel V_{ts} \ver$ (and $\vel V_{td}/V_{ts} \ver$)
}\\

\vspace*{1cm}

{\bf T. M. Aliev $^{a,}$\footnote{taliev@cc.emu.edu.tr}},~~ 
{\bf C. S. Kim $^{b,}$\footnote{kim@cskim.yonsei.ac.kr,~ 
http://phya.yonsei.ac.kr/\~{}cskim/}}~~and~~
{\bf M. Savc{\i} $^{c,}$\footnote{savci@newton.physics.metu.edu.tr}}\\

\vspace*{0.5cm}

 $a:$ Physics Department, Girne American University, Girne, Cyprus \\
 $b:$ Physics Department, Yonsei University, Seoul 120-749, Korea \\
 $c:$ Physics, Department, Middle East Technical University, 06531
      Ankara, Turkey \\
       
\vspace{1.0cm}
 
\vspace*{1.5cm}

\end{center}        

\begin{abstract}
\vspace{0.5cm}

\noindent
We examine the possibility for precise determination
of $\vel V_{ts} \ver$ (and $\vel V_{td}/V_{ts} \ver$) 
from the exclusive decays, 
$B \rar K \ell^+ \ell^-,~ B \rar K^* \ell^+ \ell^-$ (and 
$B \rar \pi \ell^+ \ell^-,~ B \rar \rho \ell^+ \ell^-$).
We show that the ratio $\vel V_{ts} \ver$
can be extracted experimentally with a small theoretical uncertainty
from  hadronic form--factors, 
if we {\bf appropriately constrain } kinematical regions of $q^2$.
We also give detailed analytical and numerical results on 
the differential decay width $d\Gamma(B \rar K^* \ell^+ \ell^-)/dq^2$, 
and the ratios of integrated branching fractions,
${\cal B}(B \rar \rho \ell^+ \ell^-)/{\cal B}(B \rar K^* \ell^+ \ell^-)$.
We estimate that one can determine the ratio $\vel V_{ts} \ver$ 
from those decays within theoretical accuracy of $\sim 10\%$.

\end{abstract}

\vspace{1cm}

\newpage

\section{Introduction}

The determination of the elements of the Cabibbo--Kobayashi--Maskawa
(CKM) matrix is one of the most important issues of quark flavor
physics. The precise determination of $V_{td}$ and $V_{ub}$ elements
is of special significance, since they are closely related to the  
origin of CP violation in the Standard Model (SM).  
Furthermore, the accurate knowledge of these
matrix elements can be useful in relating them to the fermion masses
and also in searches for hints of new physics beyond the SM.
For these reasons many strategies for the accurate determination of 
$V_{td}$ and $V_{ub}$ are under intensive investigation \cite{R1}. 
The main reason
we are interested in $B$ physics is that this area is very likely to
yield information about new physics beyond the SM. 
We expect that new physics will influence 
experimentally measurable quantities in different ways. For example,
most of us expect that $\Delta B=2$ transitions are more sensitive
to new physics than decay rates. New physics may couple
differently to $K$ mesons compared to $B$ mesons. 
Therefore, it is essential
to determine the CKM matrix elements in as many different methods
as possible.

In the existing literature, 
we can find several proposals of different methods for precise 
determination of $V_{td}$ and/or $|{V_{td} / V_{ts}}|$ \cite{R1,R2,R3,R4}:
\begin{itemize}
\item
$|V_{td}|$ can be extracted indirectly through
$B_d - \overline{B}_d$ mixing.
However, in $ B_d -\overline{B}_d$ mixing 
the large uncertainty of the hadronic
matrix elements prevents us from extracting CKM elements with
good accuracy.  
\item
A better extraction of $|{V_{td} / V_{ts}}|$ can be made
if $ B_s -\overline{B}_s$ is measured as well, because
the ratio $f_{_{B_d}}^2 B_{_{B_d}}/f_{_{B_s}}^2 B_{_{B_s}}$
can be predicted much more reliably.
\item
The determination of $|{V_{td} / V_{ts}}|$ from the ratios
of rates of several hadronic two--body $B$ decays, such as  
$\Gamma(B^0 \to \overline{K}^{*0} K^0)
   /\Gamma(B^0 \to \phi K^0)$, 
$\Gamma(B^0 \to \overline{K}^{*0} K^{*0})
   /\Gamma(B^0 \to \phi K^{*0})$, 
$\Gamma(B^+ \to \overline{K}^{*0} K^+) 
   /\Gamma(B^+ \to \phi K^+)$, and 
$\Gamma(B^+ \to \overline{K}^{*0} K^{*+})
   /\Gamma(B^+ \to \phi K^{*+})$ 
has also been proposed in \cite{R2}.
\item
$V_{td}$ can be determined from $K \to\pi \nu\bar\nu$, 
$B \to\pi \nu\bar\nu$ and $B \to \rho \nu\bar\nu$
decays with small theoretical uncertainty \cite{R1,R3}.
\item
In \cite{R4} a new method was proposed for the determination of 
$\vel V_{td} / V_{ts} \ver$ from the ratio of the inclusive decay 
distributions
\bea
{\ds{\frac{d{\cal B}}{ds}}(B \rar X_d \ell^+ \ell^-)}/
{\ds{\frac{d{\cal B}}{ds}}(B \rar X_s \ell^+ \ell^-)}~,\nnb
\eea
where $s$ is the dilepton invariant mass.
\end{itemize}

In this work, in order to determine $ \vel V_{ts} \ver $
(and the ratio $ \vel V_{td} /V_{ts} \ver $ ) 
we carefully examine another well known method
from an analysis of exclusive decay distributions
\begin{displaymath}
{\ds{\frac{d{\cal B}}{ds}}}(B \rar \pi \ell^+ \ell^-),~~
{\ds{\frac{d{\cal B}}{ds}}}(B \rar K \ell^+ \ell^-),~~ 
{\ds{\frac{d{\cal B}}{ds}}}(B \rar \rho \ell^+ \ell^-)
~~ {\rm and}~~~
{\ds{\frac{d{\cal B}}{ds}}}(B \rar K^* \ell^+ \ell^-).
\end{displaymath}
It is well known that the experimental investigation and
detection of exclusive decays are much easier 
than those of inclusive ones, although 
the theoretical understanding of exclusive decays is complicated
considerably by nonperturbative hadronic form factors. 
Exclusive $B \rar M \ell^+ \ell^-$ decays have been previously studied
in \cite{R5,R6} in the framework of the heavy quark effective theory  
\cite{R7}. Later these decays were also examined for new physics 
effect  \cite{R8}.
As is well known, the investigation of the rare decays 
$B \rightarrow \ell^+ \ell^-$ in future $B$ factories, KEK-B, SLAC-B,
B-TeV and LHC-B, would provide us of one of the best way to
determine $|V_{ts}|$ (and $|V_{td}/V_{td}|$).

This paper is organized as follows. In Section 2 we present
analytic expressions for 
$ {d \Gamma}/{d s}(B \rar M \ell^+ \ell^-),~ M = \pi,\rho,K,K^*$.
In Section 3 we study numerical distribution of 
$ {d \Gamma}/{d s}(B \rar K^* \ell^+ \ell^-)$ and
the ratios of branching fractions
\begin{displaymath}
 {\cal B}(B \rar \rho \ell^+ \ell^- ) /
 {\cal B}(B \rar K^* \ell^+ \ell^- )~~~ {\rm and}~~~
 {\cal B}(B \rar \pi \ell^+ \ell^- ) / 
 {\cal B}(B \rar K \ell^+ \ell^- ) .
\end{displaymath}

\section{Theory of  $B \rar M \ell^+ \ell^-$ ($M=\pi,K,\rho,K^*$) decays}

In the Standard Model the process $B \rar M \ell^+ \ell^-$ 
$(M = \pi,~K,~\rho,~ K^*) $ is described at quark level 
by $b \rar q \ell^+ \ell^-$ $(q = s,~d)$
transitions and receives contributions from $Z$ and $\gamma$ mediated
penguins and box diagrams.. The QCD corrected Hamiltonian for 
$b \rar q \ell^+ \ell^-$ decay can be written \cite{R9,R10,R11,R12} as 
\bea 
{\cal H}_{eff} = - \frac{4 G_F}{\sqrt{2}} V_{tq}^* V_{tb}
\sum_{i=1}^{10} C_i {\cal O}_i +
\frac{4 G_F}{\sqrt{2}} V_{uq}^* V_{ub} \Big[ C_1 \ga {\cal O}_1^u - 
{\cal O}_1 \dr + C_2 \ga {\cal O}_2^u - {\cal O}_2 \dr \Big]~,
\eea
where $V_{ij}$ are the CKM matrix elements.
The operators are given as
\bea
{\cal O}_1 &=& \ga {\bar q}_{L\alpha} \gamma_\mu b_{L\alpha} \dr 
        \ga {\bar c}_{L\beta} \gamma^\mu c_{L\beta} \dr~, \nnb \\ 
{\cal O}_2 &=& \ga {\bar q}_{L\alpha} \gamma_\mu b_{L\beta} \dr
       \ga {\bar c}_{L\beta} \gamma^\mu c_{L\alpha} \dr~, \nnb \\ 
{\cal O}_3 &=& \ga {\bar q}_{L\alpha} \gamma_\mu b_{L\alpha} \dr
\sum_{q^\prime = u,d,s,c,b} 
\ga {\bar q^\prime}_{L\beta} \gamma^\mu q^\prime_{L\beta} \dr~, \nnb \\
{\cal O}_4 &=& \ga {\bar q}_{L\alpha} \gamma_\mu b_{L\beta} \dr
\sum_{q^\prime = u,d,s,c,b} 
\ga {\bar q^\prime}_{L\beta} \gamma^\mu q^\prime_{L\alpha} \dr~, \nnb \\
{\cal O}_5 &=& \ga {\bar q}_{L\alpha} \gamma_\mu b_{L\alpha} \dr
\sum_{q^\prime = u,d,s,c,b}
\ga {\bar q^\prime}_{R_\beta} \gamma^\mu q^\prime_{R_\beta} \dr~, \nnb \\
{\cal O}_6 &=& \ga {\bar q}_{L\alpha} \gamma_\mu b_{L\beta} \dr
\sum_{q^\prime = u,d,s,c,b} 
\ga {\bar q^\prime}_{R_\beta} \gamma^\mu q^\prime_{R_\alpha} \dr~, \nnb \\
{\cal O}_7 &=& \frac{e}{16 \pi^2} {\bar q}_\alpha \sigma_{\mu\nu} 
\ga m_b R + m_q L \dr b_\alpha F^{\mu\nu}~, \nnb \\
{\cal O}_8 &=& \frac{g}{16 \pi^2} 
{\bar q}_\alpha  T^a _{\alpha\beta}\sigma_{\mu\nu} 
\ga m_b R + m_q L \dr b_\beta G^{a\mu\nu}~, \nnb \\
{\cal O}_9 &=&  \frac{e^2}{16 \pi^2} \ga {\bar q}_\alpha \gamma^\mu L 
b_\alpha \dr \ga \bar \ell \gamma_\mu \ell \dr~, \nnb \\
{\cal O}_{10} &=& \frac{e^2}{16 \pi^2} \ga {\bar q}_\alpha \gamma^\mu L 
b_\alpha \dr \ga \bar \ell \gamma_\mu \gamma_5 \ell \dr~, \nnb \\ \nnb \\
{\cal O}_1^u &=& \ga  {\bar q}_{L\alpha} \gamma_\mu b_{L\alpha} \dr
\ga  {\bar u}_{L\beta} \gamma^\mu u_{L\beta} \dr~, \nnb \\ 
{\cal O}_2^u &=& \ga  {\bar q}_{L\alpha} \gamma_\mu b_{L\beta} \dr  
\ga  {\bar u}_{L\beta} \gamma^\mu u_{L\alpha} \dr~, \nnb
\eea
where $L(R) = \frac{1}{2} \ga 1 \pm \gamma_5 \dr$ are the chiral projection
operators. 

Using the effective Hamiltonian in Eq.(1), the resulting QCD corrected
matrix element for the decays $b \rar q \ell^+ \ell^-$ $(q=d,~s)$ can be
written as 
\bea
{\cal M} &=& \frac{G_F \alpha}{2 \sqrt{2} \pi} V_{tq}^* V_{tb}\, 
\Bigg\{ C_{9_q}^{eff} \, \bar q \gamma_\mu \ga 1 - \gamma_5 \dr 
b \, \bar \ell \gamma_\mu \ell + C_{10} \, \bar q \gamma_\mu 
\ga 1 - \gamma_5 \dr b \, \bar \ell \gamma_\mu\gamma_5 \ell  \nnb \\
&-& 2 C_7^{eff} \, \bar q \, i \sigma_{\mu\nu} \frac{q^\nu}{q^2} 
\left[ m_q \ga 1 - \gamma_5 \dr + m_b \ga 1 + \gamma_5 \dr \right] 
b \, \bar \ell \gamma_\mu \ell \Bigg\}~,
\eea
where 
\bea
C_{9_q}^{eff} = \tilde C_9 + 
Y_{LD}^q \ga \hat s \dr~.
\eea
In Eq. (2),  $q^\nu$ is the four momentum transfer to dileptons,
$\hat s = q^2/m_b^2$ and 
\bea 
\tilde C_9 = C_9 \, \Bigg\{ 1 + \frac{\alpha_s \ga \mu \dr }{\pi}
\omega \ga \hat s \dr \Bigg\} + Y_{SD}^q \ga \hat s \dr~.
\eea
The function $Y_{SD}^q\ga \hat s \dr$ is the one--loop matrix element of 
${\cal O}_9$ and $Y_{LD}^q\ga \hat s \dr$ 
describes the long distance contributions 
due to the vector $ J/\psi,~\psi^\prime,~\cdots$ resonances. The function
$\omega \ga \hat s \dr$ represents the one gluon correction to the matrix
element of the operators ${\cal O}_9$. 
Its explicit form can be found in \cite{R9,R12}. 
The explicit forms of the two functions 
$Y_{SD}^q\ga \hat s \dr$ 
and $Y_{LD}^q\ga \hat s \dr$ are given \cite{R4} as
\bea
Y_{SD}^q\ga \hat s \dr &=& g \ga \hat m_c,\hat s \dr 
\left[3 C_1 + C_2 + 3 C_3 + C_4 + 3 C_5 + C_6 \right] \nnb \\
&-& \frac{1}{2} g \ga 1,\hat s \dr 
\left[4 C_3 +4 C_4 + 3 C_5 + C_6 \right] \nnb \\
&-& \frac{1}{2} g \ga 0,\hat s \dr
\left[ C_3 + 3  C_4 \right] 
+ \frac{2}{9} \left[ 3 C_3 + C_4 + 3 C_5 + C_6 \right] \nnb \\
&-& \frac{V_{uq}^* V_{ub}}{V_{tq}^* V_{tb}}
\left[ 3 C_1 + C_2 \right] \left[ g \ga 0,\hat s \dr - 
g \ga \hat m_c,\hat s \dr \right]~, \nnb \\ \nnb \\
Y_{LD}^q\ga \hat s \dr&=& \frac{3}{\alpha^2} \kappa 
\Bigg\{ - \frac{V_{cq}^* V_{cb}}{V_{tq}^* V_{tb}} \,C^{\ga 0 \dr} -
\frac{V_{uq}^* V_{ub}}{V_{tq}^* V_{tb}} 
\left[ 3 C_3 + C_4 + 3 C_5 + C_6 \right] \Bigg\} \nnb \\
&\times& \sum_{V_i = \psi \ga 1 s \dr, \cdots, \psi \ga 6 s \dr}
\ds{\frac{ \pi \Gamma \ga V_i \rar \ell^+ \ell^- \dr M_{V_i} }
{\ga M_{V_i}^2 - \hat s m_b^2 - i M_{V_i} \Gamma_{V_i} \dr }}~.
\eea
The function $g \ga \hat m_q,\hat s \dr$ arises 
from the one loop contributions
of the four quark operators 
${\cal O}_1$--${\cal O}_6$, {\it i.e.}, 
\bea
g \ga \hat m_q,\hat s \dr &=& - \frac{8}{9} {\rm ln} \hat m_q +
\frac{8}{27} + \frac{4}{9} y_q - 
\frac{2}{9} \ga 2 + y_q \dr \sqrt{\vel 1 - y_q \ver} \nnb \\
&\times& \Bigg\{ \Theta(1 - y_q) 
\ga {\rm ln} \frac{1  + \sqrt{1 - y_q}}{1  -  \sqrt{1 - y_q}} - i \pi \dr
+ \Theta(y_q - 1) 2 \,{\rm arctan} \frac{1}{\sqrt{y_q - 1}} \Bigg\}~,
\eea
where $y_q=4 \hat m_q^2/\hat s$. 
In Eq. (5), $C^{\ga 0 \dr} \equiv 3 C_1+C_2 + 3 C_3 + C_4 + 3 C_5 + C_6$.
Using $m_t=175$~GeV, $m_b=4.8$~GeV, $m_c=1.4$~GeV, 
$\alpha_s \ga m_{_W} \dr = 0.12$ and $\alpha_s \ga m_b \dr = 0.22$, 
the numerical values of the Wilson
coefficients, which we will use in our further numerical analysis, are
\bea
C_1=-0.26,~~C_2&=&1.11,~~C_3=0.01, \nnb \\
C_4=-0.03,~~C_5&=&-0.03,~~C_6=-0.03, \nnb \\
C_7=-0.32,~~C_9&=&4.26,~~{\rm and}~~C_{10}=-4.62. \nnb
\eea
The values of $C_9$ and $C_{10}$ are very
sensitive to $m_t$. We will neglect the second term in 
$Y_{LD}^q\ga \hat s \dr$, since $3 C_3 + C_4 + 3 C_5 + C_6 < C_0$. 
Masses, widths and leptonic branching ratios of the 
$J^P = 1^-$  $c \bar c$ resonances
are presented in \cite{R13}. Note that as far as short distance effects are
considered, the $u$--loop matrix element contribution to the 
$b \rar s \ell^+ \ell^-$ process is negligible due to the smallness of 
$V_{us}^* V_{ub}$ compared to $V_{cb}^* V_{cs}\simeq - V_{ts}^* V_{tb}$, 
while in the $b \rar d \ell^+ \ell^-$ case, the term proportional to
$V_{ud}^* V_{ub}$ is kept. The factor $\kappa$ is chosen to have the
value $\kappa=2.3$ \cite{R14} to reproduce the rate of decay 
$B \rar X_s J/\psi \rar X_s \ell^+ \ell^-$. 
The phase of $\kappa$ is fixed, since recent experimental data have
determined the sign of the ratio of factorization approach parameters
$a_2/a_1$ and the phase of $a_1$ is expected to be near its perturbative 
value \cite{R15}.

At this point, there arises the problem of computing the matrix elements 
of  Eq. (2) between the mesons $B$ and 
$M~\ga M=\pi,~K,~\rho,~K^*\dr$ states. The matrix element 
$\la M \ve {\cal M} \ve B \ra$ has been investigated 
in the framework of different
approaches, such as chiral perturbation
theory \cite{R16}, three point QCD sum rules
\cite{R17}, relativistic quark model \cite{R18}, effective heavy quark
theory \cite{R5}, and light cone QCD sum rules \cite{R19}--\cite{R22}.

The matrix elements for $B \rar P \ell^+ \ell^-$ $(P = \pi,~K)$ decays 
can be written in terms of the form--factors 
\bea 
\la P ( p_2 ) \vel \bar q \gamma_\mu ( 1 - \gamma_5) b \ver B (
p_1 ) \ra &=& 
f_+ ( q^2 ) ( p_1 + p_2 )_\mu
+ f_-( q^2 ) q_\mu ~, \nnb \\ \nnb \\
\la P ( p_2 ) \vel \bar q i \sigma_{\mu\nu} q^\nu b \ver B (p_1 ) \ra &=&
\left[ ( p_1 + p_2 )_\mu q^2 - (m_B^2 - m_P^2) q_\mu \right]
\frac{f_T(q^2)}{m_B + m_P}~,
\eea
where $q=p_1-p_2$.

The matrix elements for $B \rar V \ell^+ \ell^-~(V=\rho,~K^*)$ decays 
are defined as follows
\bea
\lefteqn{
\la V (p_2, \epsilon) \vel \bar q \gamma_{\mu}( 1- \gamma_5) b \ver B(p_1)
\ra = } \nnb \\
&-&  \epsilon_{\mu \nu \alpha \beta} \epsilon^{*\nu} p_2^\alpha q^\beta
\frac{2 V(q^2)}{m_B + m_V} -
i \epsilon_\mu^* ( m_B + m_V) A_1(q^2)
+ i  (p_1 + p_2)_\mu (\epsilon^* q) \frac{A_2(q^2)}{m_B + m_V} \nnb \\
&+& i  q_\mu (\epsilon^* q) \frac{2 m_V}{q^2} \left[ A_3(q^2) - A_0 (q^2)
\right]
~,\\ \nnb \\ 
\lefteqn{
\la V (p_2, \epsilon) \vel \bar q i \sigma_{\mu \nu} q^\nu (1+ \gamma_5) b
\ver B(p_1) \ra = } \nnb \\
&& 4 \epsilon_{\mu \nu \alpha \beta} \epsilon^{* \nu} p_2^\alpha
q^\beta T_1 (q^2)
+  2 i \left[ \epsilon_\mu^* (m_B^2 - m_V^2) - (p_1 + p_2)_\mu (\epsilon^*
q)
\right] T_2 (q^2) + \nnb \\
&+&  2 i (\epsilon^* q) \left[ q_\mu - (p_1+p_2)_\mu \frac{q^2}{m_B^2 -
m_V^2}
 \right] T_3 (q^2)~,
\eea   
where $\epsilon$ is the 4--polarization vector of the $V$--meson. Using the
equation of motion, the form--factor $A_3(q^2)$ can be written as a linear
combination of the form--factors $A_1(q^2)$ and $A_2(q^2)$ (see \cite{R17}).
\bea
A_3(q^2) = \frac{m_B + m_V}{2 m_V} A_1(q^2) - \frac{m_B - m_V}{2
m_V} A_2(q^2)~, \nnb 
\eea
with the condition $A_3(q^2=0)=A_0(q^2=0)$.

Using Eqs. (2), (7), (8), and (9) and summing over the final
lepton polarization for $B \rar P \ell^+ \ell^-$ and 
$B \rar V \ell^+ \ell^-$ decay widths, we get
\bea
\frac{d \Gamma}{d s}(B \rar P \ell^+ \ell^-) &=&
\frac{G^2 \alpha^2 m_B^5}
{2^8 3\pi^5}\vel V_{tq} V_{tb}^* \ver^2 \lambda^{3/2} \nnb \\
&\times&\Bigg\{
\vel 2 m_b C_7 \ga - \frac{f_T(q^2)}{m_B+m_P} \dr + C_9^{eff} f_+(q^2)
\ver^2 + \vel C_{10} f_+(q^2) \ver^2 \Bigg\}~,
\eea
\bea
\lefteqn{
\frac{d \Gamma}{d s}(B \rar V \ell^+ \ell^-) =
\frac{G^2 \alpha^2 m_B^3}{2^{12} 3 \pi^5} \vel V_{tq} V_{tb}^* \ver^2
s \lambda^{1/2} }\nnb \\
&&\times \Bigg\{16 \lambda m_B^4 \left[ \vel A \ver ^2
+ \vel C \ver ^2\right]
+ 2 \left[ \vel B_1 \ver ^2 + \vel D_1 \ver ^2\right]
\frac{\lambda + 12 r s}{r s}
+ 2 \left[ \vel B_2 \ver ^2 + \vel D_2 \ver
^2\right] \frac{m_B^4 \lambda^2}{r s} \nnb \\
&&- 4 \left[ {\rm Re} \ga B_1 B_2^* \dr
+ {\rm Re} \ga D_1 D_2^* \dr \right]
\frac{m_B^2 \lambda}{r s} \Bigg\}~, 
\eea
where $\lambda = 1+r^2+s^2 -2 r - 2 s - 2 r s$, $r =m_M^2/m_B^2$,
$s=q^2/m_B^2$. In Eq. (11) $A,~B_1,~B_2,~C,~D_1$ and $D_2$ are defined as
\bea 
A &=& C_9^{eff} \frac{V(q^2)}{m_B + m_V} + 4 C_7 
\frac{m_b}{q^2} T_1(q^2)~, \nnb \\ \nnb \\
B_1 &=& C_9^{eff} (m_B + m_V) A_1(q^2) + 4 C_7 \frac{m_b}{q^2} (m_B^2 -
m_V^2) T_2(q^2)~,  \nnb \\ \nnb \\ 
B_2 &=& C_9^{eff} \frac{A_2(q^2)}{m_B + m_V} 
+ 4 C_7 \frac{m_b}{q^2} \ga T_2(q^2)
+\frac{q^2}{m_B^2 - m_V^2} T_3(q^2) \dr~,  \nnb \\ \nnb \\
C &=& C_{10} \frac{V(q^2)}{m_B + m_V}~,  \nnb \\ \nnb \\
D_1 &=& C_{10} (m_B + m_V) A_1(q^2)~,  \nnb \\ \nnb \\  
D_2 &=& C_{10} \frac{A_2(q^2)}{m_B + m_V}~. \nnb
\eea

\section{Numerical analysis and discussions}

Now we consider the differential decay widths, 
${d{\Gamma}}/{dq^2}(B\rightarrow (\pi,\rho,K,K^*) + \ell^+ + \ell^- )$.
The hadronic formfactors in the framework of  light-cone QCD sum rules
have been calculated in  \cite{R19}--\cite{R22}.
For those values of the formfactors, we have used the results of
\cite{R19,R20}, where  the radiative corrections to the leading twist 
contribution and $SU(3)$ breaking effects are also taken into account.
The $q^2$ dependence of the formfactors can be represented in terms of 
three parameters as
\bea
F(q^2) = \frac{F(0)}{1-a_F\,\frac{q^2}{m_B^2} + b_F \left 
    ( \frac{q^2}{m_B^2} \right)^2}~, \nnb
\eea
where the values of parameters $F(0)$, $a_F$ and $b_F$ for the relevant
decays, $B \rar \pi$, $B \rar K$, $B \rar \rho$ and $B \rar K^*$,
are listed in Table 1.
We note that light-cone QCD sum rules method is applicable for those
decays in the region of $m_b^2 - q^2 =$ few GeV$^2$, and we found that
sum rules works very well up to  $q^2 =$ 20 GeV$^2$. 
In order to extend our investigation to the full physical phase space, 
we use the above mentioned parametrization in such a way that up to 
$q^2 =$ 20 GeV$^2$ it successfully reproduces the light-cone QCD sum
rules predictions.

\begin{table}[h]
\renewcommand{\arraystretch}{1.5}
\addtolength{\arraycolsep}{3pt}
$$
\begin{array}{|l|ccc|ccc|l|}
\hline
& F(0) & a_F & b_F & F(0) & a_F & b_F & \\ \hline
A_1^{B \rar\rho} & \phantom{-}0.26 \pm 0.04& 0.29 & -0.415& 
\phantom{-}0.34 \pm 0.05 & 0.60 & -0.023& A_1^{B \rar K^*} \\
A_2^{B \rar\rho} & \phantom{-}0.22 \pm 0.03& 0.93 & -0.092 & 
0.28 \pm 0.04 & 1.18 & \phantom{-}0.281 & A_2^{B \rar K^*}\\
V^{B \rar\rho} & \phantom{-}0.34 \pm 0.05& 1.37 & \phantom{-}0.315
 & \phantom{-}0.46 \pm 0.07 & 1.55 & \phantom{-}0.575 & V^{B \rar K^*} \\ 
T_1^{B \rar\rho} & \phantom{-}0.15 \pm 0.02 & 1.41 & \phantom{-}0.361
 & \phantom{-}0.19 \pm 0.03 & 1.59 & \phantom{-}0.615 & T_1^{B \rar K^*}\\
T_2^{B \rar\rho} & \phantom{-}0.15 \pm 0.02 & 0.28 & -0.500
 & \phantom{-}0.19 \pm 0.03 & 0.49 & -0.241 & T_2^{B \rar K^*}\\
T_3^{B \rar\rho} & \phantom{-}0.10\pm 0.02 & 1.06 & -0.076
 & \phantom{-}0.13 \pm 0.02 & 1.20 & \phantom{-}0.098 & T_3^{B \rar K^*}\\
\hline\hline
f_+^{B \rar \pi }& \phantom{-}0.30 \pm 0.04 & 1.35 & \phantom{-}0.270 &
\phantom{-}0.35 \pm 0.05 & 1.37 &  \phantom{-}0.350
& f_+^{B \rar K}\\
f_T^{B \rar \pi }& -0.30 \pm 0.04 & 1.34 & \phantom{-}0.260 & 
-0.39 \pm 0.05 & 1.37 & \phantom{-}0.370& f_T^{B \rar K} \\ \hline  
\end{array}
$$
\caption{$B$ meson decay form factors in a three-parameter fit, where the
radiative corrections to the leading twist contribution and SU(3) breaking
effects are taken into account (This Table is taken from \cite{R19,R20}).}
\renewcommand{\arraystretch}{1}
\addtolength{\arraycolsep}{-3pt}
\end{table} 

A few words about error analysis in the differential decay rates and
their ratios are in order. In both cases the errors which come from
different form--factors are added quadratically since they are 
theoretically independent of each other. Note also that all errors, 
which come from the uncertainties of the $b$ quark mass,
the Borel parameter variation, wave functions, non--inclusion of higher
twists and radiative corrections, are added in quadrature.
The uncertainty in the ratio
is estimated as the half distance between the maximum and minimum
values of the ratio.

In regard to the determination of $\vel V_{ts} \ver$, 
we first consider $B \rar K^* \ell^+ \ell^-$.
Since the expression for $B \rar V \ell^+ \ell^-$ decays 
contains many form--factors
(see Eq. (11)), each with its own uncertainty, the error in the
differential decay width may be substantial. To reduce these uncertainties,
it would be better to choose a kinematical region, 
in which the contributions from most of 
these form--factors will be practically negligible.
Therefore, we propose to
consider the end--point region, 16.5 GeV$^2 < q^2 <$ 19.25 GeV$^2$. 
In this region
the contributions from the terms that are proportional to $\sim C_7$
are much smaller than those from 
terms proportional to $\sim C_9$ or $\sim C_{10}$.
This is due to the fact that the terms which are proportional to $\sim C_7$  
contain $1/q^2$ factors. In this region the decay width takes the following
form
\bea
\frac{d \Gamma}{d s} &=&
\frac{G^2 \alpha^2 m_B^5}{2^{11} 3 \pi^5} \vel
V_{tq} V_{tb}^* \ver^2
s \lambda^{1/2} \nnb \\
&\times&
\left[ \vel \tilde C_9 \ver^2 + \vel C_{10} \ver^2 \right]
\ga 1 + \sqrt{r} \dr^2 \vel A_1 \ver^2 \frac{\lambda + 12 r s}{r}
(1 + \delta) (1 + \Delta) (1+d)~,
\eea
where $\delta$ represents the contributions from the form--factors $V$ and
$A_2$. $\Delta$ takes into account the magnetic momentum operator
contributions $(\sim {\cal O}_7)$, and $d(q^2)$ parametrizes the long
distance effects. In the above--mentioned region, the long distance
contributions are taken into account in the following way:
\bea 
\vel C^{eff}_9 \ver^2 + \vel C_{10} \ver^2 \simeq
\left[\vel \tilde C_9 \ver^2 + \vel C_{10} \ver^2 \right] 
\left[ 1 + d(q^2) \right]~. \nnb
\eea

In order to estimate the uncertainties which arise from
$d(q^2),~\Delta(q^2)$ and $\delta(q^2)$, 
we present their
$q^2$ dependences in Figs. 1, 2 and 3, respectively. 
In Fig. 1 we take into account all six
$J^{PC} = 1^{--}$ $c \bar c$ resonance states. 
From this figure we observe that the long
distance contribution is about $\pm 0.18$ of the short distance
contribution. As is obvious from Fig. 2,
the contribution of the magnetic dipole operator is about
$ \Delta \simeq -0.13 \pm 0.01$, whose behavior is observed to be almost 
independent of $q^2$.
The same contribution calculated in the framework of
the HQET gives numbers quite close to our result, in the range   
$-0.18 < \Delta < -0.14$ \cite{R13}, in the same region. 
In Fig. 3, the dependence of $\delta(q^2)$ on $q^2$ is depicted. As is clear
from this figure, the contribution of $\delta$ to the differential decay
width is substantial at lower values of $q^2$. However, the error due
to the uncertainties in the form--factors only causes about 
$\sim 1\%$ deviation from the
case where central values of all form--factors are used.
In Fig. 4 we present the dependence of 
$d\Gamma(B \rar K^* \ell^+ \ell^-)/dq^2$ on $q^2$ and indeed, 
with the observed oscillatory behavior,
the uncertainty that 
$d(q^2)$ brings to the $B \rar K^* \ell^+ \ell^-$ differential decay rate is 
about $\sim 8 \%$, although its leading uncertainty to the perturbative
contribution is about $\pm 0.18$. 

We note that the extraction of  $|V_{td}/V_{ts}|$  
from the decay widths 
$B \rar \rho \ell^+ \ell^-$ and $B \rar K^* \ell^+ \ell^-$            
in the low invariant mass region, for example 
2 GeV$^2 <q^2 <m_{J/\psi}^2$,
becomes more  problematic. 
In this region the contribution of the magnetic
dipole operator ${\cal O}_7$ is large. Therefore the theoretical predictions
of the decay widths of the processes $B \rar \rho \ell^+ \ell^-$ and 
$B \rar K^* \ell^+ \ell^-$ 
have significant uncertainties in this region, 
since the form--factors $T_1,~T_2$ and $T_3$ play
an essential role. In our opinion, it is better to 
investigate the $B \rar \pi \ell^+ \ell^-$ and $B \rar K \ell^+\ell^-$ 
decays in this low momentum region for the determination
of the ratio $\vel V_{td}/V_{ts} \ver$, 
since both decays are described only by
the two form--factors $f_+$ and $f_T$ (when lepton masses are neglected).
Various theoretical schemes
predict that $f_+(0)\simeq - f_T(0)$ (see \cite{R17}--\cite{R22}). In this
region of $q^2$, that is far from $q^2_{max}$, the $B^*$ pole contribution
\cite{R14} is unlikely to violate the SU(3) symmetry predictions. 

We now show the possibility for 
extracting $\vel V_{td}/V_{ts} \ver$ from the ratio
\bea 
\frac{d {\rm R}}{d q^2} \equiv  \frac{d}{dq^2}
\left[ {{\cal B} (B \rar \rho  \ell^+ \ell^-)}/
{{\cal B} (B \rar K^* \ell^+ \ell^-)} \right]~, 
\eea
and the ratio 
\bea
\frac{d {\rm R}_1}{dq^2} \equiv \frac{d}{dq^2}
\left[ {{\cal B} (B \rar \pi  \ell^+ \ell^-)}/
{{\cal B} (B \rar K \ell^+ \ell^-)} \right] 
\simeq \vel \frac{V_{td}}{V_{ts}} \ver^2 
\vel \frac{f_+^{B \rar \pi}(q^2)}{f_+^{B \rar K}(q^2)} \ver^2~.
\eea
Performing the integrations over $q^2$, 
from 16.5 GeV$^2$ to 19.25 GeV$^2$ in
Eq. (13) and from 2 GeV$^2$ to 9 GeV$^2$ in Eq. (14), i.e.
\bea
{\rm R}(16.5 < q^2 < 19.25~{\rm GeV}^2) &=& 
\frac{{\cal B} (B \rar \rho  \ell^+ \ell^-)}
{{\cal B} (B \rar K^* \ell^+ \ell^-)} ~, \nnb \\
{\rm R}_1(2 < q^2 < 9~{\rm GeV}^2) &=& 
\frac{{\cal B} (B \rar \pi  \ell^+ \ell^-)}
{{\cal B} (B \rar K \ell^+ \ell^-)} ~, \nnb
\eea 
we get the numerical results 
\bea
{\rm R} &=& 0.85 \pm 0.06~, \nnb \\
{\rm R}_1 &=& 0.77 \pm 0.06~, \nnb 
\eea 
(normalized to $\vel V_{td}/V_{ts} \ver^2$).

At this point we estimate the number of expected events
of the types $B \rar V \ell^+ \ell^-$ and $B \rar P \ell^+ \ell^-$ in
experiments at future $B$ factories. Future symmetric and
asymmetric $B$ factories of electronic and hadronic colliders should 
produce much more than $10^9$ $B$--$\bar B$ mesons per year by the year 2010. 
Assuming $10^9$ $B$ mesons effectively reconstructed, the number of expected
events in the corresponding kinematical regions
(16.5~GeV$^2 < q^2 <$ 19.25~GeV$^2$ for $B \rar V \ell^+ \ell^-$ decay and
2~GeV$^2 < q^2 < $ 9~GeV$^2$ for $B \rar P \ell^+ \ell^-$ decay)
are calculated to be
\bea
{\rm N}(B \rar K^* \ell^+ \ell^-)&=&{\cal B}(B \rar K^* \ell^+ \ell^-) 
\times 10^9 \sim ~140,\nnb \\
{\rm N}(B \rar K \ell^+ \ell^-)&=&{\cal B}(B \rar K \ell^+ \ell^-)
\times 10^9 \sim ~735,\nnb \\
{\rm N}(B \rar \rho  \ell^+ \ell^-)&=&{\cal B}(B \rar \rho \ell^+ \ell^-) 
\times 10^9 \sim ~5,\nnb \\
{\rm N}(B \rar \pi  \ell^+ \ell^-)&=&{\cal B}(B \rar \pi  \ell^+ \ell^-)
\times 10^9 \sim ~28.\nnb
\eea
As can be seen easily, for determination of $|V_{td}|$ or 
$|V_{td}/V_{ts}|$ we need much more than 10$^9$ $B$ mesons produced,
and it will be only possible at future hadronic $B$ factories, such as
B-TeV and LHC-B, where the double lepton triggering  helps high 
reconstruction
efficiencies with more than 10$^{11}$ $B$ mesons per year produced.
From the above results we conclude that 
these decays have a good chance to be detected at future $B$ factories.

The problem of accurate determination of CKM matrix elements with different
methods receives special attention in connection with the fact that the
next generation of $B$ meson decay experiments will be a test of the flavor
sector of the SM at high precision as well as allowing the determination of 
$V_{td}$ and $V_{ub}$ with very high accuracy.
Simultaneous determinations of CKM angles and phases 
\cite{R1,R23} would be extremely important to check 
the consistence within the SM 
and to search for the hints of new physics  beyond the SM.
It is well known that the experimental investigation and
detection of exclusive decays is much easier 
than those of inclusive ones, while 
the theoretical understanding of exclusive decays is much more complicated
due to nonperturbative hadronic form factors. 
We have found that for the determinations of $\vel V_{ts} \ver$
(and $\vel V_{td}/V_{ts} \ver$), 
due to the character of the relevant hadronic form factors,
the kinematic region of small $q^2$ is more useful for
the $B \rar P  \ell^+ \ell^-$ decays, 
while the kinematical region of $q^2$ near the end--point is more suitable
for the $B \rar V  \ell^+ \ell^-$ decays. And we estimated 
numerically $\vel V_{ts} \ver$ and the ratio
$\vel V_{td}/V_{ts} \ver$ from the ratios of those decays with 
a theoretical uncertainty of  $\sim 10\%$ .

\vspace*{1cm}

\noindent
{\Large \bf Acknowledgments}
\\

\noindent
We thank M. Drees and T. Morozumi for the careful reading of the 
manuscript and their valuable comments.  
The work of T.A. and M.S. was supported by TUBITAK.
C.S.K. wishes to acknowledge the financial support of Korea Research 
Foundation made in the program year of 1997.

\newpage

\section*{Figure Captions}
{\bf 1.} Dependence of $d$ on $q^2$ for $B \rar K^* \ell^+\ell^-$
decay. Dash--dotted $d(q^2)=0$ line corresponds 
to the short distance contributions.\\ \\
{\bf 2.} Dependence of $\Delta$ on $q^2$ for $B \rar K^* \ell^+\ell^-$
decay. Dotted and dash--dotted curves correspond to the cases when
the uncertainty is added and subtracted from the central values of all
form--factors, respectively.\\ \\ 
{\bf 3.} Dependence of $\delta$ on $q^2$ for $B \rar K^* \ell^+\ell^-$
decay. Dotted and dash--dotted curves correspond to the cases when the
uncertainty is added and subtracted from the central values of all
form--factors, respectively. \\ \\
{\bf 4.}  Dependence of $d\Gamma(B \rar K^* \ell^+\ell^-)/dq^2$ on $q^2$
in units of
\bea
\frac{G^2 \alpha^2 m_B^3}{2^{11} 3 \pi^5} \vel V_{ts} V_{tb}^* \ver^2
\left[ \vel \tilde C_9 \ver^2 + \vel C_{10} \ver^2 \right]
\ga 1 + \sqrt{r} \dr^2~. \nnb 
\eea
In this figure the wavy curve numbered 
as 2 represents the case that takes into account all
$1^{--}$ states of $c\bar c$ resonances for the central values of 
all form--factors,
while the wavy curves numbered as 1 and 3, correspond to the cases 
when the uncertainty is
added and subtracted from the central values of all
form--factors, respectively.
The dash--dotted line represents
the contribution of only the three lightest resonances. 
The dotted curve, on the other hand, is for
the perturbative result, i.e., $d(q^2)=0$.

\newpage

\begin{figure}[tb]
\vspace*{-5cm}
\hspace*{-2cm}
\centerline{\epsfig{figure=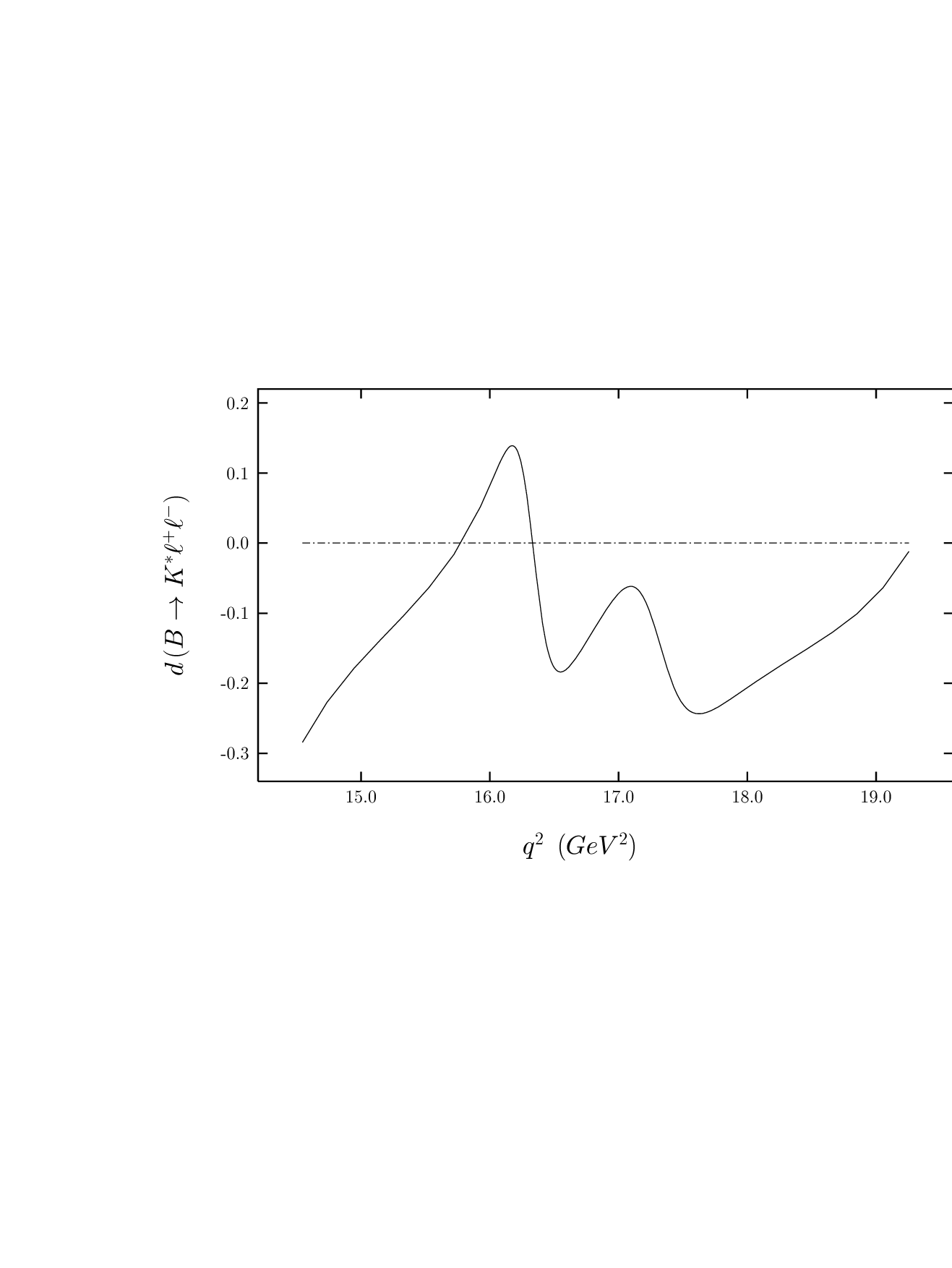,height=18cm,width=17cm,angle=0}}
\vspace*{-5cm}
\caption{  }
\end{figure}

\begin{figure}[tb]
\vspace*{-5cm}
\hspace*{-2cm}
\centerline{\epsfig{figure=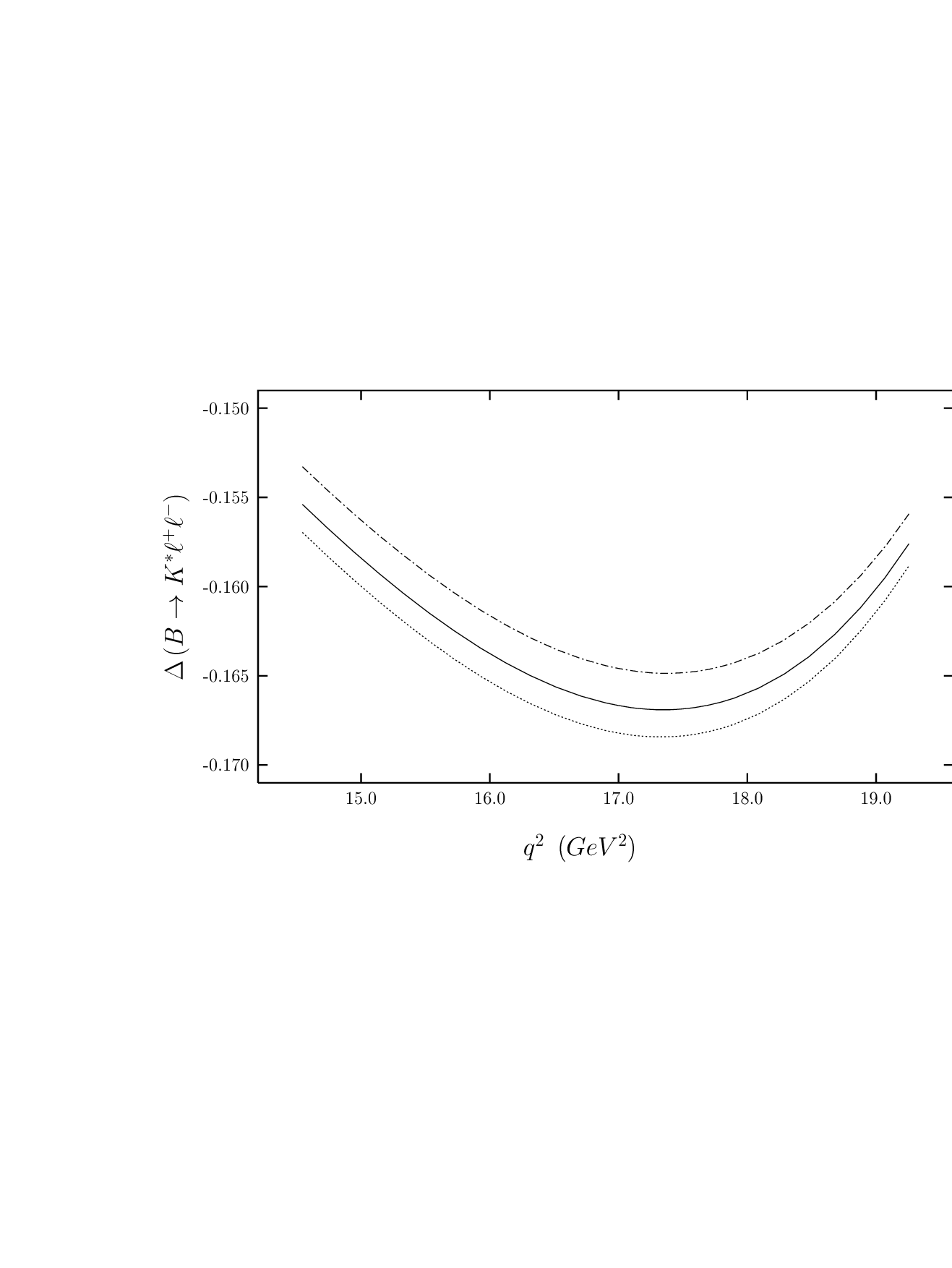,height=18cm,width=17cm,angle=0}}
\vspace*{-5cm}
\caption{  }
\end{figure}

\begin{figure}[tb]
\vspace*{-5cm}
\hspace*{-2cm}  
\centerline{\epsfig{figure=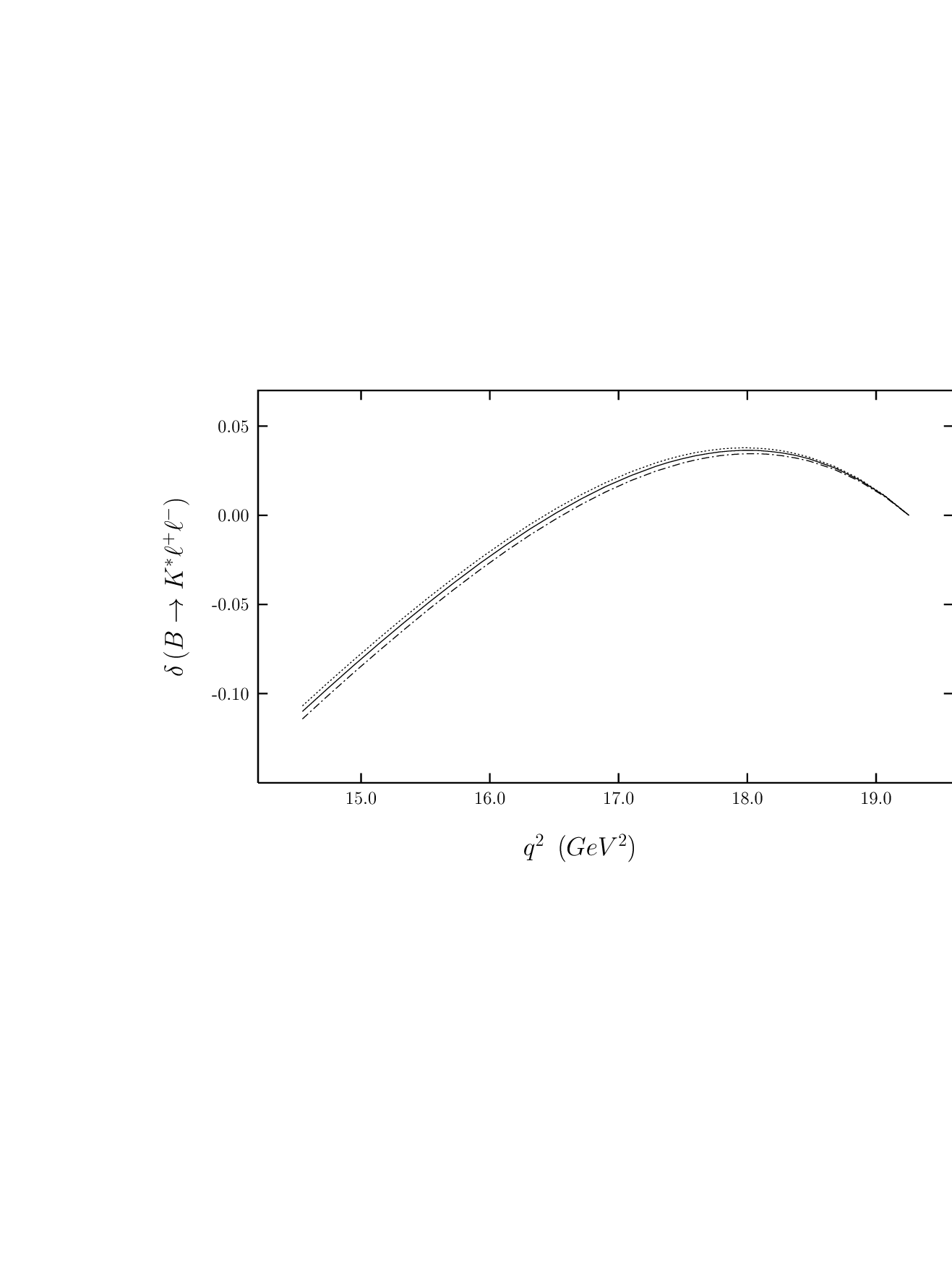,height=18cm,width=17cm,angle=0}}
\vspace*{-5cm}
\caption{  }
\end{figure}

\begin{figure}[tb]
\vspace*{-5cm}
\hspace*{-2cm}
\centerline{\epsfig{figure=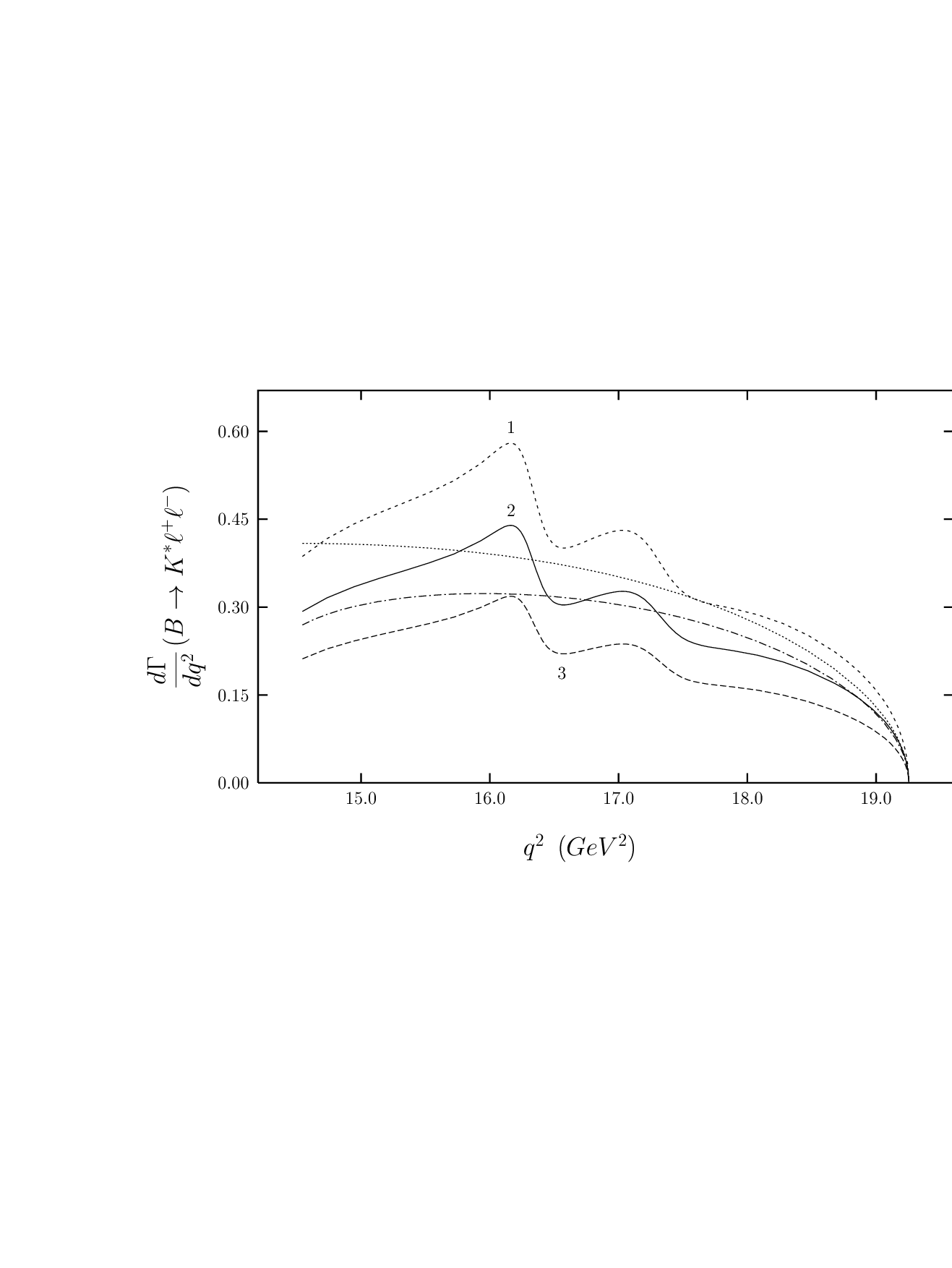,height=18cm,width=17cm,angle=0}}
\vspace*{-5cm}
\caption{  }
\end{figure}

\end{document}